%Paper: hep-th/9404116
%From: DASMAHAPATRA S <S.Dasmahapatra@city.ac.uk>
%Date: Tue, 19 Apr 1994 15:29:28 +0100 (BST)

\input harvmac
\nref\rBethe{H. A. Bethe, Z. Phys. 71 (1931), 205.}
\nref\rLieb{E. Lieb, Phys. Rev. 130 (1963), 1616.}
\nref\ryangs{C. N. Yang and C. P. Yang, J. Math. Phys. 10 (1969), 1115.}
\nref\rABF{G. E. Andrews, R. J. Baxter and P. J. Forrester, J. Stat. Phys. 35
(1984) 193.}
\nref\rBRone{V.~V. Bazhanov and N.~Yu. Reshetikhin,
 Int. J. Mod. Phys. A4 (1989) 115.}
\nref\rBRtwo{V.~V. Bazhanov and N.~Yu. Reshetikhin, J.~Phys.~A23 (1990), 1477.}
\nref\rTak{M.~Takahashi, Prog. Theo. Phy. 46 (1971), 401.}
\nref\rstd{M. Takahashi and M. Suzuki, 48 (1972), 2187, M. Gaudin, Phys. Rev.
Lett., 26 (1971), 1301.}
\nref\rBab{H.~M. Babujian, Nucl. Phys. B 215, (1983), 317.}
\nref\rADMone{G. Albertini, S. Dasmahapatra and B.~M. McCoy, Int. J.
 Mod. Phys. A7, Suppl. 1A (1992) 1.}
\nref\rJKMO{M. Jimbo, A. Kuniba, T. Miwa and M. Okado, Comm. Math. Phys., 119
(1988), 543.}
\nref\rBRthree{V.~V. Bazhanov and N.~Yu. Reshetikhin, Prog. Theor. Phys. Suppl.
102 (1990) 301.}
\nref\rKuniba{A. Kuniba, Nucl. Phys. B389 (1993) 209.}
\nref\rADMtwo{G. Albertini, S. Dasmahapatra and B.~M. McCoy, Phys. Lett. A
 170 (1992) 397.}

\nref\rKedMc{R. Kedem and B.~M. McCoy, J. Stat. Phys. 71, (1993), 865.}
\nref\rDKMM{S. Dasmahapatra, R. Kedem, B.~M. McCoy and E. Melzer, J. Stat.
Phys. (in press), hep-th/9304150.}
\nref\rKKMMone{R.Kedem, T.~R. Klassen, B.~M. McCoy, and E. Melzer,
 Phys. Lett. B 304, 263.}
\nref\rKKMMtwo{R.Kedem, T.~R. Klassen, B.~M. McCoy, and E. Melzer,
 Phys. Lett. B 307, (1993), 68.}
\nref\rDKKMM{S. Dasmahapatra, R. Kedem, T.~R. Klassen, B.~M. McCoy and
E. Melzer in Yang-Baxter Equations in Paris, World Sci., also in Int. J.
Mod. Phys. B, Vol. 7, (1993), 3617.}
\nref\rRS{B. Richmond and G. Szekeres, J. Austral. Soc. (Series A) 31
(1981), 362.}
\nref\rNRT{W. Nahm, A. Recknagel, M, Terhoeven, Mod. Phys. Lett. A, 8 (1993),
1835.}
\nref\rTer{M. Terhoeven, Bonn preprint, hep-th.}
\nref\rKNS{A. Kuniba, T. Nakanishi, J. Suzuki, Harvard preprint HUPT-92/A069.}
\nref\rstring{S. Dasmahapatra, ICTP preprint IC/93/91,  hep-th/9305024.}

\nref\rHal{F.~D.~M. Haldane, Phys. Rev. Lett. 61, 937 (1991).}
\nref\rtrieste{S. Dasmahapatra, to be published in the proceedings
of the summer school and workshop on string theory held in Trieste, 1993.}
\nref\rGKO{P. Goddard, A. Kent and D. Olive, Comm. Math. Phys., 103 (1986),
105.}
\nref\rLP{J. Lepowsky and M. Primc, {\it Structure of standard modules for the
affine Lie algebra} $A_1^{(1)}$, Cont. Math., Vol. 46 (AMS, Providence, 1985).}
\nref\rFeigin{B. Feigin and A. Stoyanovsky, preprint, RIMS-942, 1993.}
\nref\rBerk{A. Berkovich, preprint, BONN-HE-94-04, hep-th/9403073.}
\nref\rFatZam{V. A. Fateev and A. B. Zamolodchikov, Phys. Lett. A 92 (1982),
37.}

\nref\rwilf{H. Wilf in ``Surveys in Combinatorics, 1989,'' ed. J.
Siemons, LMS Lec. Not. Ser. 141, Cambridge, 1989.}
\nref\rg{G. Albertini, J. Phys. A 25 (1991), 1799.}
\nref\rAl{G. Albertini, preprint, SB-ITP-93-64, hep-th/9310133 (to appear
in Int. J. Mod. Phys. A).}
\nref\rNien{V.~V. Bazhanov, B. Nienhuis and S.~O. Warnaar, Phys. Lett.B 322
(1994), 198.}
\nref\ramp{G. Albertini, B.M. McCoy, J.H.H. Perk, Eigenvalue spectrum
of the superintegrable chiral Potts model, Adv. in Pure Math. 19
(1989) 1.}
\nref\rdkm{S. Dasmahapatra, R. Kedem and B.M. McCoy, Spectrum and
completeness of the three-state superintegrable chiral Potts model,
Nucl. Phys. B396 (1993) 506.}

\nref\rGN{F.~M. Goodman and T. Nakanishi, Phys. Lett. B262 (1991), 259.}

\nref\rKirillov{A.~N. Kirillov, preprint, hep-th/9312084.}
\nref\rKircount{A.~N. Kirillov, J. Sov. Math., 36 (1987).}
\nref\rComb{S. Kerov, A.~N. Kirillov and N.~Yu. Reshetikhin, J. Sov. Math.
41 (1988), 916, A.~N. Kirillov and N.~Yu. Reshetikhin, J. Sov. Math. 41
(1988), 925, A.~N. Kirillov and N.~Yu. Reshetikhin, Lett. Math. Phys. 12,
(1986), 500.}

%%%%%%%%%%%%%%
\Title{\vbox{\baselineskip12pt\hbox{City University Preprint CMPS 94-103}
\hbox{hep-th/yymmnn}}}
{\vbox{\centerline{On State Counting and Characters}}}

\vskip 10mm
\centerline{ \rm Srinandan Dasmahapatra}
\vskip 10mm
\centerline{\it Department of Mathematics }
\centerline{\it City University}
\centerline{\it Northampton Square, London EC1V OHB, U.K.}

\vskip 20mm
\centerline{\bf Abstract}
\vskip 5mm

We outline the relationship between the thermodynamic densities and
quasi-particle descriptions of spectra of RSOS models with an underlying
Bethe equation. We use this to prove completeness of states in some
cases and then give an algorithm for the construction of branching
functions of their emergent conformal field theories. Starting from
the Bethe equations of $D_n$ type, we discuss some aspects of the $Z_n$
lattice models.

\Date{\hfill 4/94}
\vfill\eject

\newsec{Introduction}

A number of questions of interest regarding interacting
many-body quantum systems are posed within the framework of the limit
of the number of
particles going to infinity. Two situations in which will be relevant to
the discussion in this paper are the thermodynamics at non-zero
temperatures and the low-energy eigenvalue spectrum of the hamiltonian.
To set up the thermodynamic formalism, it is imperative that one has a
control over counting the states of the system, in order to compute the
entropy. On the other hand, the low-energy spectrum spectrum is,
generically, of a form that
is interpreted as that describing``quasi-particles," which are then said
to populate the allowed energy levels subject to composition rules. In
order to determine these rules, it is again necessary to classify and
count the states.

The main point of this paper is twofold -- first, we point out a direct
(and innocent-looking) connection between
the thermodynamic formalism and the quasi-particle description in a wide
class of one-dimensional quantum spin systems whose spectral information
is expressed in terms of solutions to the Bethe equations. This will
then enable us to carry out a detailed counting of the states on a finite
lattice, and consequently to contruct the
branching functions of the emergent conformal field theories.

A quick word about where to place these results: The correspondence between
2-dimensional lattice models of statistical mechanics, their 1-dimensional
spin-chain counterparts and their field theory limits -- both conformal
(massless) and their integrable off-critical (massive) counterparts, have
long been explored in the context of the particular models we consider. In
\rBRone,
the study of the critical generalized RSOS models was undertaken using
"Bethe ansatz techniques,"  that had earlier been developed  for
excitation spectra \rLieb and for thermodynamics  \ryangs, (see also \rTak
\rstd).
In addition, the authors of \rBRone computed the central charges
by taking the $T\rightarrow 0$ limit of the thermodynamic calculation of
the entropy, following \rBab.  These methods have been further extended for
the critical models of higher rank algebras (\rBRtwo, \rKuniba) and for
the off-critical models whose masses have been determined and
S-matrices computed in \rBRthree.
The key ingredient in all of these studies is the densities of the roots
of the Bethe equations, and to quote the authors of \rBRthree,
"the set of densities ... determines all macroscopical observables in the model
(pg. 307)."

In \rKedMc, \rDKMM, the composition rules (\rADMone) and the quasi-particle
spectrum (\rADMtwo) were used to construct the partition function of
the low-temperature quantum 3-state Potts spin chain, which were shown
to to give $q$-series expansions of known modular branching functions of
the associated conformal field theories. The work of \rNRT used the method
of \rRS to extract the central charge by taking the $q\rightarrow 1^{-}$
limit of the $q$-series, uncovering, in the process, the form of the
dilogarithm identities very similar to those that had featured in the
central charge computations via the thermodynamics of these
Bethe ansatz solvable models.  This similarity was exploited to formulate
new $q$-series identities for branching functions in \rKKMMone , \rKKMMtwo,
\rTer and \rKNS. The work of \rstring served as a bridge between the
thermodynamic calculations of \rBRone, \rBRtwo and \rKuniba and the
quasi-particle approach of \rKedMc and \rDKMM, thus ``explaining" away
the origin of these identities into the structure of the Bethe equations
themselves. The key step that allowed this connection is the innocuous
equation (4.4). The proofs of some of these forms were already known (\rLP),
but since then, proofs of some more of these identities have appeared
(\rFeigin, \rBerk).

A brief outline of this paper: Section 2 explicitly shows the counting
procedure that was followed in \rADMone to get the composition rules for
the simplest series of RSOS models. Section 3 outlines the procedure for
those models based on the simply laced algebras. Section 4 sets up the
correspondence between the thermodynamic and the quasi-particle
descriptions.  Section 5 outlines the algorithm for setting up the
$q$-series that count the low-lying states which are (conjectured, in
general, to be) branching functions in the vacuum sector of the coset
conformal field theory.  Section 6 talks about the $Z_n$ models
of \rFatZam as an example, and the Appendix works out a useful combinatoric
identity to count the states in the ground state sector of arbitrary spin
$su(2)$ models.

\newsec{Counting states in the ABF models}

The RSOS models that were introduced in \rABF.
In \rBRone, the eigenvalues of the transfer matrix of these models were
expressed in terms of the roots of the following equation (in their notation,
$l=r-2$):

\eqn\betheabf{\Biggl[{\sinh\bigl({\pi\over{2(l+2)}}(\lambda_j+i)\bigr)
\over\sinh\bigl({\pi\over{2(l+2)}}(\lambda_j-i)\bigr)}\Biggr]^N=\Omega_j
\prod_{k=1}^{N/2}{\sinh\bigl({\pi\over{2(l+2)}}(\lambda_j-\lambda_k+2
i)\bigr)
\over\sinh\bigl({\pi\over{2(l+2)}}(\lambda_j-\lambda_k-2 i)\bigr)}.}
$\Omega_j$ is a phase factor, $N$ the size of the chain and $l$ is an
integer.

The solutions to these equations are assumed to be of the ``string''
form \rBethe and following \rBRone, we write them in the form
\eqn\str{\lambda_j =\lambda + i (j+1-2j_1), ~~~1\leq j_1\leq j,
{}~~~1\leq j\leq l, ~~\Im m\lambda=0 .}
These assumptions were made in the bulk in \rBRone , (with a number of
caveats referring to the appearance of solutions not of the above form)
but here we shall impose these on finite-size lattices. (Of course,
the same thermodynamic limit can be recovered.)
On multiplying out the equations for the components of each string,
\rTak \rstd we
get a set of equations for the real parts, $\lambda$, of the strings.
We then take the logarithm of \betheabf  so that the integer branches are
distinct:
\eqn\log{N t_{j}(\lambda_{\mu}^{j })~=~2 \pi i I_{j \mu} +
\sum_{k=1}^{l} \sum_{\nu=1}^{M_k} \Theta_{j k}
 (\lambda_{\mu}^{j}-\lambda_{\nu}^{k}),}
where $\lambda_{\mu}^{j}$ labels the center of the $\mu$-th string of
length $j$ which is a root of eq. \betheabf. The functions
$t_{j}$ and $\Theta_{j k}$ are defined below.
\eqn\tj{t_{j} (\lambda)=f(\lambda;|j-1|+1),}
\eqn\theta{\eqalign{\Theta_{j k} (\lambda)&= \Biggl(
f(\lambda;|j-k|)+2\sum_{i=1}^{min(j,k)-1} f(\lambda;|j-k|+2i) +
f(\lambda;j+k)\Biggr)
}}
\eqn\f{f(\lambda;n)=~{1\over2\pi i}\ln\Bigl({\sinh {1\over2}{\pi\over (l+2)}
(i n -\lambda)\over \sinh {1\over2}{\pi\over (l+2)} (i n +\lambda)}\Bigr), }
for integer values of $n/(l+2)$, and is $0$ otherwise.

For pedagogical reasons, we shall carry out in some detail, the
counting procudure
which is central to all further physical elaborations.  We define
\eqn\zfun{Z_j(\lambda)\equiv t_{j}(\lambda) - {1\over N}
\sum_{k=1}^{l} \sum_{\nu=1}^{M_k} \Theta_{j k}
(\lambda, \lambda_{\nu}^{k }),}
so that the (half-) integers $I_{j \mu}$ satisfy
\eqn\zrts{Z_j(\lambda_{\mu}^{j})=2\pi i{I_{j \mu}\over N}.}
If we assume that $Z_j(\lambda)$ is monotonic, then the range of the
integers, $\Delta I_j\equiv I_{j,max}-I_{j,min}$ is set by taking the
difference of the limiting
values, $Z_j(\pm\infty)$.  A result used repeatedly is the following:
\eqn\d{d_n\equiv{1\over2\pi i}[f(+\infty, n)-f(-\infty,n)] =
{l+2-n\over l+2}, \hbox{\hskip 12pt} n\leq l+1.}
Therefore,
\eqn\master{\Delta I_j=N d_j - \sum_{k=1}^l M_k\bigl\{ (1-\delta_{j
k}) d_{|j-k|} + d_{j+k} + 2\sum_{i=1}^{min(j,k)-1}
d_{|j-k|+2i}\bigr\},}
where $min(i,j)$ picks out the smaller of the two values $i$ and $j$.
Substitution \d into \master and performing the sum over the
variable $i$ takes the form
\eqn\mastera{\eqalign{(l+2)\Delta I_j& = N (l+2-j) - \sum_{k=1}^l M_k \bigl\{
\bigl( (l+2)- |j-k| \bigr) -\delta_{j k} \bigl( (l+2)-|j-k|\bigr)\cr
&\quad+\bigl( (l+2) - (j+k)\bigr) + 2 \bigl( (l+2) - |j-k|\bigr)
\bigl(min(j,k) -1 \bigr)\cr
&\quad-2 min(j,k)\bigl( min(j,k) -1\bigr)\bigr\}}.}
Splitting up the sum over $k$ to get past the $min(j,k)$ function, we
get
\eqn\masterb{\eqalign{(l+2)\Delta I_j&=N\bigl( (l+2) -j\bigr) -
\sum_{k=1}^{j-1} M_k\bigl\{ \bigl( (l+2) -j +k\bigr) + \bigl(
(l+2)-j-k\bigr) \cr
&+2 \bigl( (l+2) - j +k)(k-1) -2k (k-1)\bigr\}\cr
&-M_j\bigl\{\bigl( (l+2) -2j\bigr)-2j (j-1) + 2(l+2)(j-1) \bigr\}\cr
&-\sum_{k=j+1}^l M_k\bigl\{ \bigl( (l+2) -k +j) + \bigl(
(l+2)-j-k\bigr)\cr
& +2 \bigl( (l+2) - k +j)(j-1) -2j (j-1)\bigr\}\cr
&=N\bigl( (l+2) -j\bigr) -\bigl( (l+2) -j\bigr) \sum_{k=1}^{j-1} k M_k
+ 2j^2 M_j + (l+2)M_j\cr
& -2(l+2)j M_j - 2j \sum_{k=j+1}^l \bigl( (l+2) - k\bigr) M_k.}}

Consider the special case of the strings of length $l$: setting $j=l$
above,
\eqn\masterl{\eqalign{(l+2)\Delta I_l&=N(l+2-l) -
2(l-l+2)\sum_{k=1}^{l-1} k M_k + \bigl( 2l^2 +(l+2) -2(l+2)l \bigr)
M_l  \cr
&=2N -4\sum_{k=1}^{l-1} k M_k + M_l (2-3l),\cr
&=4({N\over2}-\sum_{k=1}^l k M_k) + 4lM_l +M_l (2-3l),\cr
&=(l+2)M_l,}}
where we have used the sum rule on the total number of roots in the last
step. We thus notice the
remarkable feature that is enforced by the string hypothesis and the
monotonicity assumptions on $Z(\lambda)$ that $\Delta I_l=M_l$, i.e.,
all the allowed values for the integers that specify the locations of
the real parts of strings of length $l$ are always occupied -- there
are ``no holes'' in this sector \rBRone.
Going back to \masterb, we replace the $M_l$ occurring in the last sum
by ${N\over2l}\sum_{k=1}^{l-1} k M_k$, and after a few
simplifications and dividing by $(l+2)$, we get
\eqn\masterc{\eqalign{\Delta I_j &= N(1-{j\over l}) + {2\over l} \sum_{k=1}^j k
M_k (j-l) + {2j\over l}\sum_{k=j+1}^{l-1} M_k (k-l) + M_j,\cr
&=N(1-{j\over l}) + 2\sum_{k=1}^{l-1} \bigl\{ min (j,k) - {jk\over
l}\bigr\}M_k\cr
&=N\bigl\{ min (j,1) - {j\cdot 1\over l}\bigr\} + 2\sum_{k=1}^{l-1}
\bigl\{ min (j,k) - {jk\over l}\bigr\}M_k.}}
Note that the coefficients in the sum in the last line in \masterc are
the ${j,k}$th matrix elements of the inverse Cartan matrix of the
$A_{l-1}$ root system!  We write the coefficient of $N$ in a
suggestive way for comparison with later results.

These integer specifications can be used to label the eigenvalues of
the hamiltonian.  Does this give a complete classification?  We shall
evaluate the number of ways these integers can be ascribed to the
$\lambda$ s that determine the eigenvalues, i.e., perform the following
cominatorial sum:
\eqn\count{ S=\sum_{\{M_j\}} \prod_{j} {\Delta I_j\atopwithdelims ()
M_j}.}
To do this we shall follow \rTak.  First of all, we reinstate $M_l$
into \count by replacing $M_1$ by the sum rule, re-expressing it in
the variables $M_2, \ldots, M_l$
\eqn\countsutwo{S=\sum_{\{M_j\}} {N/2 + \sum_{k\geq3}^l
(k-2) M_k \atopwithdelims () N/2 - \sum_{k\geq2}^l k M_k} \prod_{j=2}^{l-1}
{M_j + 2\sum_{k=j+1}^l (k-j) M_k \atopwithdelims () M_j},}
where we notice that $M_2$ now occurs in only two places, and the sum over
$M_2$ is of the form
\eqn\summtwo{ \sum_{j\geq0} {C \atopwithdelims () A-2j}
{B+j \atopwithdelims () j}} which can be evaluated as the coefficient of
$x^A$ in  $(1+x)^C (1-x^2)^{-B-1}$.

We are now left with the following:
\eqn\summthree{\sum_{\{M_k\},k\geq3} {1\over 2\pi}\oint dx
\Bigl({1+x \over x}\Bigr)^{N/2}
{1\over (1-x^2)} \prod_{k=3}^l \Bigl({ { x^k (1+x)^{k-2} }\over
(1-x^2)^{2(k-2)} }\Bigr)^{M_k} {\Delta I_k\atopwithdelims () M_k},}
where the contour is around the origin.
The sum over $M_3$ is now of the form $\sum_{j\geq0}{B+j \atopwithdelims
()j} y^j$, which is just $(1-y)^{-B-1}$. We can carry on doing this
for
successive values of $k$. In \rTak, this iterated sum has been encoded
in the following form:
\eqn\takahashi{S={1\over 2\pi}\oint dx
\Bigl({1+x \over x}\Bigr)^{N/2}{1\over 1-x^2} \prod_{k=3}^l
{1\over 1-u_j^{-1}},}
where
\eqn\recur{(u_j-1)^2=u_{j-1}u_{j+1}, \hbox{\hskip 20pt} u_3={x^3\over
(1+x)(1-x^2)}, \hbox{\hskip 20pt} u_2=x^{-2}.}
These are essentially Chebyshev polynomials of the second kind, {\it i.e.}
$u_j(x)=U_j^2({1\over2}\sqrt{1+x\over x})$, where
\eqn\chebyshev{U_n(\cos \phi )={\sin\bigl((n+1)\phi \bigr)\over
\sin\phi }.}

By looking at the poles and residues of $U_l(z)/U_{l+1}(z)$, we
notice that
\eqn\ratcheb{{U_l(z)\over U_{l+1}(z)} ={1\over l+2}\sum_{j=1}^{l+1}
{\sin^2({\pi j\over l+2})\over {z-\cos ({\pi j\over l+2})}}.}
Therefore we can express the sum as
\eqn\sumfinal{S={4\over l+2}\sum_{j=1}^{[(l+1)/2]} {1\over 2\pi}\oint dx
\Bigl({1+x \over x}\Bigr)^{N/2} {\sin^2({\pi j\over l+2})\over
{1-(4\cos^2 ({\pi j\over l+2}) -1)x}}}
(notice that the square roots cancel out). By deforming the contour to
pick up all the other poles except at the origin gives us
\eqn\sumanswer{S={2\over l+2}\sum_{j=1}^{l+1} \sin^2 ({\pi j\over l+2})
(2\cos({\pi j\over l+2}))^N,}
which is precisely the multiplicity of the singlet in the $N$-fold tensor
product of the fundamental representation of $U_q (su(2))$ at
$q=$ exp(${2\pi i \over l+2}$). This is the expected number because of
the the equivalence \rGN of the construction of the model state space
to truncated tensor product representations of quantum groups at roots
of unity. (This method of performing the sum, which is lifted from \rTak,
where it was used for the
$l\rightarrow\infty$ case, was deployeded for the $l=4$ case in \rADMone.
See also \rKirillov.)

\newsec{For the simply-laced algebras.}
A bit of history:
In \rBRone, higher spin representations of $su(2)$ were studied.
{}From the $su(2)$ models, we can generalize even further, by
considering the models of \rJKMO as solved in \rBRtwo, by starting
with the corresponding Bethe equations.  In fact, the authors of
\rBRtwo made the observation that the structure of the Bethe equations
could be cast in a way that could be generalized to the case of all
simply laced algebras even though the models whose spectrum these
equations would parametrize were not known.  This game was
further extended to the elliptic case in \rBRthree and by Kuniba \rKuniba
who set up the thermodynamics of
(hypothetical) systems whose energy eigenvalues were parametrised by
solutions to the (trigonometric) Bethe equations associated with all
({\it i.e.} not just simply-laced) untwisted affine algebras.

We shall briefly outline the counting procedure for the simply-laced
cases which closely follows section 1. This will highlight the key
feature characteristic of the connection between the quasi-particle
description (which is a recasting of the map between the integers and
momenta in a suggestive language that posits this physical paradigm)
and the thermodynamic formalism.

For the simply-laced algebras, the Bethe equations are of the form
\rBRtwo :
\eqn\betheade{\Biggl[{\sinh\bigl({\pi\over2 L}(\lambda_j^{(a)}+is
\delta_{a p})
\bigr)\over\sinh\bigl({\pi\over2 L}(\lambda_j^{(a)}-is \delta_{a p})
\bigr)}\Biggr]^N=\Omega_j^{(a)}\prod_{b=1}^r \prod_{k=1}^{N_b}
{\sinh\bigl({\pi\over2 L}(\lambda_j^{(a)}-\lambda_k^{(b)}+i C_{ab})
\bigr)\over\sinh\bigl({\pi\over2 L}(\lambda_j^{(a)}-\lambda_k^{(a)}-i
C_{ab})\bigr)}.}
\eqn\tot{N_a=N s \bigl[C^{-1}\bigr]_{a p},}
where $N$ is the size of the lattice $C$ is the Cartan matrix,
$L=l+g$, for integer $l$, $g$ is the dual Coxeter number and $s$
characterizes the type of fusion.
In what follows, the string hypothesis (as in \rBRtwo)
can be written as:

\eqn\str{\lambda_j^{(a)}=\lambda + i (j+1-2j_1), ~~~1\leq j_1\leq j,}
$\lambda$ is real and denotes the center of the string and $j$
denotes its length, which is further assumed to satisfy $1\leq j\leq l$.
We shall therefore impose
$\sum_{j=1}^{l} j M_j^{(a)}=N_a$, where $M_j^{(a)}$ denotes the number
of strings of colour (a), length $j$ and $N_a$ is given by \tot.
As before, we multiply out the Bethe equations for the components of each
string, and end up with equations for the real parts of the roots.
We then take the logarithm of the multiplied out \betheade  so that
the integer branches are  distinct:
\eqn\log{N t_{j, s}^{(a)}(\lambda_{\mu}^{j (a)})~=~2 \pi i I_{j \mu}^{(a)} +
\sum_{b=1}^r \sum_{k=1}^{l} \sum_{\nu=1}^{M_k^{(b)}} \Theta_{j k}^{(a b)}
 (\lambda_{\mu}^{j (a)}-\lambda_{\nu}^{k (b)}),}
where $\lambda_{\mu}^{j (a)}$ labels the center of the $\mu$-th string of
length $j$ and color $(a)$ which is a root of eq. \betheade. The functions
$t_{j, s}^{(a)}$ and
$\Theta_{j k}^{(a b)}$ are defined below.
\eqn\tjs{t_{j, s}^{(a)} (\lambda)=\delta_{a p}\sum_{k=1}^{min(j,s)}
f(\lambda;|j-s|+2k-1),}
\eqn\theta{\eqalign{\Theta_{j k}^{(a b)} (\lambda)&=\delta_{a b} \Biggl(
f(\lambda;|j-k|)+2\sum_{i=1}^{min(j,k)-1} f(\lambda;|j-k|+2i) +
f(\lambda;j+k)\Biggr)
\cr&~~~~~~~-I_{a b}\Biggl(\sum_{i=1}^{min(j,k)} f(\lambda;|j-k|+2i-1)\Biggr),}}
\eqn\f{f(\lambda;n)=~{1\over2\pi i}\ln\Bigl({\sinh {1\over2}{\pi\over L}(i n
-\lambda)\over \sinh {1\over2}{\pi\over L} (i n +\lambda)}\Bigr), }
for integer values of $n/L$, and is $0$ otherwise. $I_{a b}$ is the incidence
matrix of the respective Dynkin diagrams.
%(For the non-simply laced cases, the branches are chosen following the same
%principle \rADMone, but they look a little more complicated, and for the
%case of $B_n$ and $F_4$, an extra prescription is required.)
As in section 2, we define the corresponding $Z(\lambda)$, this time with
an extra colour index.  Assuming monotonicity, we find that
for strings of length $l$, the range of integers $\Delta I_l^{(a)}$
coincides with the total number of strings of that length, $M_l^{(a)}$,
for all colour labels $(a)$. There are {\it no holes} in the distribution
of integers for these strings.
Since the $l$-strings do not contribute to
the counting of states, we can eliminate $M_l^{(a)}$ using the sum rule
\tot. The ranges of integers associated with the other strings are:
\eqn\rangem{\Delta I_j^{(a)}= N \delta_{a p} \bigl[ C_{A_{l-1}}^{-1}
\bigr]_{j s}
+ M_j^{(a)} - \sum_{b=1}^r \sum_{k=1}^{l-1} (C_{\cal G})_{a b}
\bigl[ C_{A_{l-1}}^{-1}\bigr]_{j k} M_k^{(b)},\hbox{\hskip.5cm}j<l,}
where $C_{\cal G}$ is the Cartan matrix of the (simply-laced) Lie algebra
$\cal G$ and (as before) $C_{A_{l-1}}^{-1}$ is the inverse Cartan
matrix of $A_{l-1}$. Note that this specializes to the case of $su(2)$
considered earlier, for $s=1$.
This gives a combinatorial count of the number of states of the form
described above,
\eqn\count{ \sum_{\{M_j^{(a)}\}} \prod_{j,a} {\Delta I_j^{(a)}\atopwithdelims
() M_j^{(a)}}.}

We shall use the equality of binomial coefficients
\eqn\bin{ {A\atopwithdelims () B}={A\atopwithdelims () A-B}}
to define a new variable
\eqn\defn{ N_j^{(a)}=\Delta I_j^{(a)} - M_j^{(a)}, }
which counts the number of holes. In these variables, the allowed range
of integers takes the form:
\eqn\rangen{\Delta I_j^{(a)}= N \delta_{s j} \bigl[ C_{\cal G}^{-1}
\bigr]_{a p}
+ N_j^{(a)} - \sum_{b=1}^r \sum_{k=1}^{l-1} (C_{\cal G}^{-1})_{a b}
\bigl[ C_{A_{l-1}}\bigr]_{j k} N_k^{(b)}.
}

The observation that $\Delta I_l^{(a)}=M_l^{(a)}$ had been first made in
the context of thermodynamic densities studied in \rBRone, \rBRtwo. The
striking difference is, of course, that in the thermodynamic formalism,
$N$ is a large number, and only  {\it a single} order in $N$ is retained,
and there was no reason to expect this to be true at any finite $N$. (That
the authors of \rBRone notice deviations from the string
hypothesis \str makes it all the more surprising.) We shall exploit this
coincidence in order to do away with the complication of choosing branches
as in \log.

\newsec{The thermodynamic formalism.}

The procedure of setting up the thermodynamics has now become standard,
going back to \ryangs. We shall
reproduce the equations in \rKuniba where the thermodynamics of systems
based on the Bethe equations corresponding to all simply-laced algebras
were studied.
\eqn\bethe{\Biggl[{\sinh\bigl({\pi\over2 L}(\lambda_j^{(a)}+i(s \omega_p |
\alpha_a ))
\bigr)\over\sinh\bigl({\pi\over2 L}(\lambda_j^{(a)}-i(s \omega_p | \alpha_a ))
\bigr)}\Biggr]^N=\Omega_j^{(a)}\prod_{b=1}^r \prod_{k=1}^{N_b}
{\sinh\bigl({\pi\over2 L}(\lambda_j^{(a)}-\lambda_k^{(b)}+i (\alpha_a |
\alpha_b))
\bigr)\over\sinh\bigl({\pi\over2 L}(\lambda_j^{(a)}-\lambda_k^{(a)}-i
(\alpha_a | \alpha_b))\bigr)}.}
\eqn\tot{N_a=N s \bigl[C^{-1}\bigr]_{a p},}
where $N$ is the size of the lattice $\alpha_a$ are the simple roots,
$\omega_p$ are the fundamental weights, $(\cdot | \cdot)$ is the canonical
bilinear form on the dual
to the Cartan subalgebra, $C$ is the Cartan matrix, $L=l+g$, for integer
$l$, $g$ is the dual
Coxeter number and $s$ characterizes the type of fusion.

The string hypothesis for the solutions of \bethe is
\eqn\str{\lambda_j^{(a)}=\lambda + i t_a^{-1}  (j+1-2j_1), ~~~1\leq j_1\leq j}
with $t_a^{-1}=(\alpha_a|\alpha_a)/2$. $\lambda$ is real and denotes the
center of the string and $j$
denotes its length, which is further assumed to satisfy $1\leq j\leq t_a l$.
We shall therefore impose
$\sum_{j=1}^{t_a l} j M_j^{(a)}=N_a$, where $M_j^{(a)}$ denotes the number
of strings of colour (a), length $j$ and $N_a$ is given by \tot.

To make the idea of upgrading bulk quantities to finite lattices concrete,
we define densities for strings and holes \ryangs  by requiring their
integrals over
all $\lambda$ to be exactly equal to the number of strings and holes divided
by $N$, the size of the system. This is the precise statement that rules the
interpolation between the bulk and finite lattices. In the notation of
\rKuniba,
\eqn\densities{ \hat{\rho}_j^{(a)}(0)=M_j^{(a)}/N , ~~~ \hat{\sigma}_j^{(a)}(0)
=N_j^{(a)}/N,}
where $\hat{f}(0)$ denotes the Fourier transform of $f$ for zero argument:
\eqn\fourier{\hat{f}(x)=\int_{-\infty}^{\infty} f(\lambda)e^{-i\lambda
x}d\lambda, ~~~ \hat{f}(0)=\int_{-\infty}^{\infty} f(\lambda) d\lambda.}
One would expect this correspondence to be only
true modulo correction terms vanishingly small compared to $N$.
In the thermodynamic limit, because solving the integral equations involves
taking derivatives on the densities, one can choose the branches of the
logarithms (of the Bethe equations) freely. What we see is that there exists
a choice of branch on the finite lattice that makes the translation of
statements true in the bulk to finite lattices possible.

The equations governing the densities that determine the thermodynamics
of these systems as studied in \rKuniba are
\eqn\Kuniba{\delta_{p a} \hat{A}_{p a}^{(l) s m} = \hat{\sigma}_m^{(a)} +
\sum_{b=1}^r \sum_{k=1}^{t_b l-1} \hat{M}_{a b} \hat{A}_{a b}^{(l) m k}
\hat{\rho}_k^{(b)},}
where
\eqn\defkun{\eqalign{\hat{A}_{a b}^{(J) m k}=\hat{A}_{b a}^{(J) k m} &=
\;{ \sinh\bigl( min(m/t_a, k/t_b)x\bigr) \sinh\bigl(
(J-max(m/t_a,k/t_b))x\bigr)
\over{\sinh(x/t_{a b}) \sinh(J x)}},\cr
\hat{M}_{a b}=\hat{M}_{b a}&={{t_b}\over{t_{a b}}} C_{a b}+2\delta_{a b}
\bigl(\cosh(x/t_a)-1\bigr),}}
and $t_{a b}=\hbox{\rm max}(t_a, t_b)$.

Setting $x$ (the Fourier transform variable) to zero,
and applying the definition \densities to \Kuniba, we get
\eqn\generaln{N_j^{(a)}=N \delta_{a p} [C_{A_{t_p l-1}}^{-1}]_{j s} -
\sum_{b=1}^r \sum_{k=1}^{t_b l-1} K_{a b}^{j k} M_k^{(b)},}
where
\eqn\defK{K_{a b}^{j k} \equiv (\alpha_a | \alpha_b) \{ \hbox{\rm min}(t_b j,
t_a k) - {jk\over l}\}.}
We can also solve for $M_j^{(a)}$ in terms of $N_j^{(a)}$ to get
\eqn\generalm{M_j^{(a)}=N\sum_{k=1}^{t_p l-1} (K^{-1})_{a p}^{j k}
[C_{A_{t_p l-1}}^{-1}]_{k s} - \sum_{b=1}^r \sum_{k=1}^{t_b l-1}
(K^{-1})_{a b}^{j k} N_k^{(b)},}
where $K^{-1}$ is defined by
\eqn\defKin{\sum_{c=1}^r \sum_{m=1}^{t_c l-1} (K^{-1})_{a c}^{j m}
K_{c b}^{m k}=\delta_{a b} \delta_{j k}.}
In the above, the index $t_a l$ never shows up because there are never any
holes in this sector, and the sum rule \tot  is used to eliminate the
dependence on $M_{t_a l}^{(a)}$. Note that for the simply-laced cases, $K$
factorizes into a ``level'' and a ``rank'' piece, and \generaln  reduces to
\rangem  using \defn. The range of integers for the finite system, under
this correspondence, is $\Delta I_j^{(a)}=M_j^{(a)} + N_j^{(a)}$.

\newsec{Quasi-particles and $q$-series.}

Now for some physics. The first thing to remember is that the
hamiltonians are connected to transfer matrices of two
dimensional statistical mechanical systems with positive-definite
Boltzmann weights, and for either of the $\pm$ signs the ground state
has to be unique by the Perron-Frobenius theorem. Our classification
(assumed complete)
via the partition of the roots of \bethe into stringy
clusters for the states in this sector (where the ground state is
expected to lie) indicates that the only states that meet this criterion
are given by the partition for which the summand in binomial coefficients
gives $1$.
Using the variable variable $N_j^{(a)}$
which counts the number of holes in the allowed distribution of integers,
and look for solutions of $N_j^{(a)}=0$.  The observation that the range of
allowed integers for strings of length $t_a l$ was equal to the number $M_{t_a
l}$
itself indicates that the state in which all the roots are of this type is
one such candidate for a ground state. The only other state is then the
state which has a distribution of strings given by
\eqn\ground{M_j^{(a)}=N\sum_{k=1}^{t_p l-1}
(K^{-1})_{a p}^{j k} [C_{A_{t_p l-1}}^{-1}]_{k s} .}
We thus have the two ground states
for the two hamiltonians that differ by an overall sign.  Any other partition
of the roots then corresponds to an excited state in either model.

\eqn\exm{\Delta I_j^{(a)}=N \delta_{a p} [C_{A_{t_p l-1}}^{-1}]_{j s} +
M_j^{(a)}
- \sum_{b=1}^r \sum_{k=1}^{t_b l-1} K_{a b}^{j k} M_k^{(b)},}
and
\eqn\exn{\Delta I_j^{(a)}=N\sum_{k=1}^{t_p l-1} (K^{-1})_{a p}^{j k}
[C_{A_{t_p l-1}}^{-1}]_{k s} + N_j^{(a)} - \sum_{b=1}^r
\sum_{k=1}^{t_b l-1} (K^{-1})_{a b}^{j k} N_k^{(b)}}
encode the taxonomy of all the states of the system {\it except} the ground
state which has been ``subtracted out"
We can then regard \exm and \exn as the fundamental equations
that determine the excitation spectra of the two models, with the numbers
$M_i^{(a)}$ and $N_i^{(a)}$ counting the number of these excitations.

A generically striking property that the spectrum of a many-body quantum
system (such as the ones described here) possesses is one described as
 a quasi-particle form.  This refers to the
additive decomposition of the eigenvalues thus:
\eqn\qpone{\lim_{N\rightarrow\infty} E-E_{GS}=\sum_{\alpha, rules}
\sum_j^{n_\alpha} e^{\alpha}(p^{\alpha}_j),}
where $E$ refers to the eigenvalues of the hamiltonian ($E_{GS}$ being the
smallest eigenvalue) and $N$ refers to the size of the system. The
functions in the summand are the dispersion curves of what are defined
to be  quasi-particles of different species $\alpha$, and the index
$j$ runs over  their number $n_\alpha$. The momentum of a particular
eigenstate is
\eqn\qptwo{P-P_{GS}=\sum_{\alpha, rules} \sum_j^{n_\alpha} p^{\alpha}_j.}
The ``rules'' determine the composition of the eigenstate in question,
and in particular, could depend on the statistics of the excitations.

The $n_\alpha$ quasi-particles of \qpone and \qptwo are the
excitations which number $M_j^{(a)}$ and $N_j^{(a)}$ above the two vacua
in the thermodynamic limit.  The completeness argument on the finite lattice
(which we have demonstrated only for a subset of the cases under
consideration) indicates the  completeness of the particle picture of the
(separable) Hilbert space of the emergent field theory.

The momentum eigenvalue \qptwo is written as a sum of the logarithm
of the term on the left hand sides of \bethe. Thus, the
sum of the integers for any set of roots gives the total momentum (with the
appropriate $2\pi\over N$ taken into account) of the state
that these roots correspond to. Therefore we see that for every term in
$\Delta I_j^{(a)}=M_j^{(a)}+N_j^{(a)}$ for which the term proportional to
$N$ does not have support, the contribution of these strings (or holes) to
the momentum of the eigenstate at order one ($N^0$) must necessarily vanish.
For a system with no mass gap the energy contribution must consequently be
zero to order one. However, they could (and do) contribute to the spectrum
at order $1/N$. These have been termed ``ghost'' excitations in \rDKMM.
The strings (and holes) for which $\Delta I_j^{(a)} \sim N$ have an
extensive single-excitation Hilbert space dimension and constitute the
order one excitation spectrum, and  the coefficient of $N$ encodes the
Brillouin zone scheme of these particles. (The relationship of these spectra
to their fractional statistics \rHal  has been mentioned in \rKedMc and in
\rtrieste, where the realtionship of the statistical exclusion coefficient
to the central charge of the conformal field theory has been shown.)

It is seen in numerical studies (which provided the background
for ref. \rADMone), that within each class of states with a particular root
content, the sum of the absolute values of the integers gives a good estimate
of the (approximate) degeneracy of the levels, i.e., those states with the
same value for the sum of the absolute value of the integers had energy
eigenvalues that were almost equal. This is reminiscent of the conformal
field theory definitions of energy and momentum being the sum and
differences of the eigenvalues of $L_0$ and $\bar L_0$.
One could then presume that this degeneracy would become exact in the
thermodynamic limit, and that these integers keep track of, in a rather
robust fashion and on a {\it finite} lattice, a classification of states
that take on significance (in terms of dynamical symmetries) {\it only} in
the thermodynamic limit.
We can then hope to count degenerate states  in the $N\rightarrow\infty$
limit by keeping track of these integers. (For more discussion on
quasi-particles and state counting in the thermodynamic limit, see \rDKKMM.)

In other words, the binomial counting may be refined by introducing a
grading, i.e., by introducing a variable $q$, whose powers are the
integers. Consider, for example, any $3$ of the $5$ integers in the
set $\{\pm2, \pm1, 0\}$ can be chosen such that their sum takes values
in the set $\{\pm3, \pm2, \pm1, 0 \}$. The multiplicity of their
occurence can be observed from the coefficient of the appropriate
power of $q$ in \eqn\exq{{5 \atopwithdelims \{ \}  3}_q =
q^{-3}+q^{-2}+2 q^{-1}+2 + 2q + q^2 + q^3,}
where
\eqn\qnum{ {A+B\atopwithdelims \{ \} B}=q^{-{1\over2}AB}
{A+B\atopwithdelims [] B} }
are defined in terms of $q$-binomial coefficients,
\eqn\qbin{ {A\atopwithdelims [] B} ~= ~~{(q;q)_A \over (q;q)_{A-B}
(q;q)_B},}
where
\eqn\qfac{(q;q)_A~=~\prod_{j=1}^A (1-q^j),}
(non-zero only for integers, $A$ and
$B$, with $0\leq A\leq B$).

Construct the following:
\eqn\charsa{\eqalign{f_p&=\sum_{\{M_j^{(a)}\}} \prod_{j,(a)}
q^{-{1\over2} M_j^{(a)} N_j^{(a)}(M_j^{(a)})}
{M_j^{(a)} + N_j^{(a)}(M_j^{(a)})\atopwithdelims [] M_j^{(a)}},\cr
f_h&=\sum_{\{N_j^{(a)}\}} \prod_{j,(a)}
q^{-{1\over2} N_j^{(a)} M_j^{(a)}(N_j^{(a)})}
{N_j^{(a)} + M_j^{(a)}(N_j^{(a)})\atopwithdelims [] N_j^{(a)}},
}}
where the parantheses are used to indicate that we write $f_p$ only in the
$M_j^{(a)}$ variables and $f_h$
only in the $N_j^{(a)}$ variables.
Since large values of $\lambda$ correspond to small momentum contributions
and therefore, low energy states, and the integers have a monotonic
dependence on $\lambda$, the largest integers correspond to the lowest
energy. We therefore assign zero momentum to the term in $\Delta I_j^{(a)}$
proportional to $N$ (which is the edge of the Brillouin zone)
in the power of $q$ in the factor pre-multiplying the $q$-binomial
coefficients.

We can now count the low energy states in the thermodynamic limit,
$N\rightarrow\infty$.
For the massless case, this is equivalent to computing the partition
function of the theory in the particular sector, with $q=exp(-{{2\pi
v}\over NT}$), in the limit $N\rightarrow\infty$, $T\rightarrow0$, $NT$
finite, $v$ being the speed of sound.
In the $N\rightarrow \infty$ limit, the $f_p$ and $f_h$ are re-defined
as follows
\eqn\charfin{\eqalign{\chi_p&=\lim_{N\rightarrow\infty} f_p \cr
&=\lim_{N\rightarrow\infty} \sum_{\{M_j^{(a)}\}}
q^{-{1\over2} {\underline M}\cdot K\cdot {\underline M} } \times \cr
&\qquad\times\prod_{j,a}
{{N\over2}\delta_{a p} [C_{A_{t_p l-1}}^{-1}]_{j s} +  M_j^{(a)} +
\sum_{b=1}^r
\sum_{k=1}^{t_b l -1} K_{a b}^{j k} M_k^{(b)}\atopwithdelims [] M_j^{(a)}},
\cr
\chi_h&=\lim_{N\rightarrow\infty} f_h\cr
&=\lim_{N\rightarrow\infty}\sum_{\{N_j^{(a)}\}}
q^{-{1\over2} {\underline N} \cdot (K^{-1})\cdot {\underline N}} \times \cr &
\qquad\times\prod_{j,a}
{N\sum_{k=1}^{t_p l-1} (K^{-1})_{a p}^{j k}
[C_{A_{t_p l-1}}^{-1}]_{k s} - \sum_{b=1}^r \sum_{k=1}^{t_b l-1}
(K^{-1})_{a p}^{j k} N_k^{(b)} + N_j^{(a)}\atopwithdelims [] N_j^{(a)}} ,
}}
with the condition that
\eqn\limq{\lim_{N\rightarrow\infty} {N\atopwithdelims [] m}~=~{1\over
(q;q)_m} .}

These are conjectured to be the branching functions (in the vacuum sector)
corresponding to the
conformal field theories constructed as cosets ({\it a la} GKO) \rGKO.
Some of these have been conjectured earlier (see \rDKKMM for references).
The branching functions these $q$-series expansions correspond to are
listed in \rKuniba. The central charges which can be obtained by taking
the $q\rightarrow 1$ limit as in \rRS, \rNRT  and \rKKMMtwo are also listed
in \rKuniba. Thre exists a proof of the $su(2)$ formulas in \rLP for the
$M$-type variables ($\chi_{p}$) and in \rBerk for the $N$-type variables
($\chi_{h}$).

%%%%%%%%%%%%%%%%%%%%%%%%%%%

\newsec{The $Z_n$ models and some ``Snake-Oil''.}

In \rKKMMtwo it was pointed out that the $q$-series corresponding to what
in our notation would be ${\cal G}=D_n$, $s=1$ and $p=1$ and in the $M$
variable coincided with the vacuum character of $Z_n$ parafermions (with
the colour index $(a)$ taking values upto the rank, $n$).

We have obtained
\rstring this $q$-series from a Bethe equation, without any reference to
a physical model. The Fateev-Zamolodchikov model \rFatZam in its scaling
limit is known to give rise to the parafermionic conformal field theory.
The $Z_3$ case was studied in \rADMone \rADMtwo using the formulation of
\rg, and we saw that the formulas in section 1 reproduced (modulo an
orbifolding which we shall indicate later) those results (for the $Q=0$
sector of the $3$-state Potts model).  To confirm whether this counting
procedure in terms of the roots of \betheade gives the total number of
$Q=0$ states, we need to execute a corresponding bimomial sum.

We shall number the nodes of the Dynkin diagram of $D_n$ such that the
point of bifurcation is at $3$ and the ends of the fork are at $1$ and $2$.
The expression we are going to evaluate is
$$S_M=  \sum_{ \{m_j \} }{M+m_3 \atopwithdelims () 2m_1}
{m_3 \atopwithdelims () 2m_2}{m_1+m_2+m_4 \atopwithdelims ()
2m_3}{m_3+m_5 \atopwithdelims ()2m_4} \ldots {m_{n-2}+m_n
\atopwithdelims () 2m_{n-1}}{m_{n-1} \atopwithdelims () 2m_n}.$$
Recall that for the critical $3$-state Potts case, which is the $n=3$
case of the $Z_n$ models of Fateev and Zamolodchikov, the number of
genuine $O(1)$ quasi-particle excitations were both even and odd in
number unlike the corresponding orbifold-related RSOS model which only
had an even number in keeping with the $Z_2$ invariance.  The even
sector only picks up the states with the $Z_2$ positive charge states.
We therefore expect that to recover the complete $Z_n$ charge zero
sector we will have to sum over $m_1$ half integral as well, which
will necessarily make the associated $m_2$ to be half-inegral too.

Following the elegant and extremely useful method advocated by H. Wilf
called the``Snake Oil'' method \rwilf,
we define
\eqn\fxdef{f(x)=\sum_M x^M S_M,}
interchange the order of summation and perform the sum over $M$ first.
We shall follow the convention that the binomial coefficient ${A
\atopwithdelims ()B}$ vanishes if $B<0$ or if $A$ is a positive
integer less than $B$. Using the formula (valid for $|x|<1$),
\eqn\useful{\sum_{k\geq 0}  {k \atopwithdelims ()B} x^k = {1\over
(1-x)}\bigl({x\over 1-x}\bigr)^B,}
we get
$$\sum_{M\geq 0}x^M {M+m_3 \atopwithdelims () 2m_1+\alpha} = {x^{2m_1
+\alpha-m_3} \over (1-x)^{2m_1+\alpha+1}}.$$
(Here $\alpha=0,1$ which we have introduced in order to keep track of
the odd/evenness of the number of excitations.)
Plugging this result into the sum over $m_1$, we then need to perform
$${x^{-m_3+\alpha}\over (1-x)^{\alpha+1}}\sum_{m_1}\bigl(({x\over
1-x})^2\bigr)^{m_1} {m_1+m_2+m_4+\alpha \atopwithdelims ()2m_3},$$
which is again of the form \useful.
After some simplification, the sum over $m_2$ then takes the form
\eqn\summtwo{\bigl({1-x\over x}\bigr)^{2m_4} {1-x\over 1-2x}
{x^{3m_3} \over (1-2x)^{2m_3}} \sum_{m_2} \bigl({1-x\over x})^{2 m_2
+\alpha} {m_3 \atopwithdelims () 2m_2+\alpha} .}

The presence of the $\alpha$s alongside the summed (dummy) variable
indicates that the sum is performed over all (even and odd)
the integers, and thus we have
\eqn\gensum{\eqalign{f(x)&=\sum_{m_n}\ldots\sum_{m_3} {1-x\over 1-2x}
\bigl({x\over 1-2x}\bigr)^{2m_3} \bigl({1-x\over x}\bigr)^{2m_4}
{m_3+m_5 \atopwithdelims () 2m_4}\times\cr & \qquad \qquad
{m_4+m_6 \atopwithdelims () 2m_5}\ldots
{m_{n-2}+m_n \atopwithdelims () 2m_{n-1}}{m_{n-1}
\atopwithdelims () 2m_n}}.}

Define $f_3\equiv (x/(1-2x))^2$, so that upon performing the
sum over $m_3$,
\eqn\gensuma{\eqalign{f(x)&={1-x\over 1-2x}{1\over 1-f_3}\sum_{m_n,\ldots,m_4}
f_3^{-m_5}\bigl({1-x\over x}{f_3\over1-f_3}\bigr)^{2m_4}
{m_4+m_6 \atopwithdelims () 2m_5}\times\cr
& \qquad \qquad {m_5+m_7\atopwithdelims () 2m_6} \ldots {m_{n-2}+m_n
\atopwithdelims
() 2m_{n-1}}{m_{n-1} \atopwithdelims () 2m_{n}}}.}
Defining
\eqn\defffour{f_4\equiv \bigl({1-x\over x}{f_3\over
1-f_3}\bigr)^2=\bigl({x\over 1-3x}\bigr)^2, }
and iterating this process, we end up with
\eqn\ffinaldn{f(x)={1-x\over1-2x}\prod_{j=3}^n {1\over 1-f_j}, }
with
\eqn\dnfs{(f_j^{-1} - 1)^2 = f_{j-1}^{-1}f_{j+1}^{-1}, }
and the initial two values of $f_j$ for $j=3, 4$ are defined above. It
can be checked that a solution to these recurrence relations with the
given initial data is given by
\eqn\defsoldn{f_j=\bigl({x\over1-(j-1)x}\bigr)^2.}
Plugging it all into \ffinaldn, we get
\eqn\ansdn{f(x)={1-(n-1)x\over 1-nx}=\sum n^{M-1} x^M,}
which precisely counts the number of states in the ($Z_n$ charge) $Q=0$
sector of the Fateev-Zamolodchikov spin chain.

Going back to the characters, we now consider those in the $N$ variables,
which we now expect are those that describe the conformal field theory
of the anti-ferromagnetic chain. In \rKKMMtwo and \rTer, it was conjectured
(based on expansions of the formulas \charfin as $q$-series, using
{$Mathematica^{TM}$}) that these characters are those
at radii $\sqrt{n\over2}$ on the Gaussian line of the moduli space of
$c=1$ theories. This expectation has been borne out in \rAl for odd $n$.

\newsec{Discussion.}

In order to describe the physics that the solutions to the Bethe equations
encoded, we invoked the positivity of the Boltzmann weights of {\it
non-existent} models.  For the anti-ferromagnetic
$D_n$ models, the classification scheme seemed to describe the states
of the $Z_n$ models of \rFatZam whose Boltzmann weights were, however, not
positive for the anti-ferromagnetic regime (of the associated spin chain).
This condition of ``confinement of holes" served to
rearrange the classification of states in a useful way giving correct
results. (In passing, let us note that in \rNien, the $E_8$ structure of the
Bethe equations emerges owing to a similar occurence of a `` `frozen'
Dirac sea.")

The more interesting questions that remain to be answered involve cases
where the dependence of the energy functional on the roots of the Bethe
equations is non-trivial, as in the case of the superintegrable chiral
Potts model \ramp, where similar completeness studies have been made
\rdkm.

\vskip 1cm
\noindent {\bf Acknowledgement}
\vskip .2cm
It is a pleasure to thank Prof. Barry M. McCoy for ongoing discussion
and encouragement and to Neil Calkin for a couple of fruitful e-mail
exchanges.

\listrefs

\newsec{Appendix}

In this appendix, we shall evaluate the sum over the products of
binomial coefficients for the case of $su(2)$, arbitrary spin $s$,
by the ``Snake-Oil'' method.
We shall warm up to it by first re-doing the sum for the $s=1$ case.
Let us write the combinatorial summand in terms of the ``hole''
variables (we shall change notation, replacing $N_j$ by $m_j$, and
$N/2\equiv M$):
\eqn\sumfund{S_M = \sum_{m_1, m_2,\ldots ,m_n} {M+m_2 \atopwithdelims () 2m_1}
{m_1+m_3 \atopwithdelims () 2m_2}{m_2+m_4 \atopwithdelims ()
2m_3}\ldots {m_{n-2}+m_n \atopwithdelims () 2m_{n-1}}{m_{n-1}
\atopwithdelims () 2m_{n}}.}
Next, define
\eqn\fxdef{f(x)=\sum_M x^M S_M,}
interchange the order of summation and perform the sum over $M$ first:

$$\sum_{M\geq 0}x^M {M+m_2 \atopwithdelims () 2m_1}=\sum_{M\geq 0}
x^{-m_2} x^{M+m_2}{M+m_2 \atopwithdelims () 2m_1}.$$
We shall follow the convention that the binomial coefficient ${A
\atopwithdelims ()B}$ vanishes if $B<0$ or if $A$ is a positive
integer less than $B$. Using the formula (valid for $|x|<1$,
\eqn\useful{\sum_{k\geq 0}  {k \atopwithdelims ()B} x^k = {1\over
(1-x)}\bigl({x\over 1-x}\bigr)^B,}
we get
$$f(x)=\sum_{m_1, m_2,\ldots ,m_n} {x^{-m_2}\over (1-x)}y_1^{m_1}
{m_1+m_3 \atopwithdelims () 2m_2}{m_2+m_4 \atopwithdelims ()
2m_3}\ldots {m_{n-2}+m_n \atopwithdelims () 2m_{n-1}}{m_{n-1}
\atopwithdelims () 2m_{n}},$$
where we have defined $y_1\equiv\bigl(x/(1-x)\bigr)^2$.
In the same way, we can now perform the sum over $m_1$ and be left
with
\eqn\nomone{\eqalign{f(x)&=\sum_{m_2,\ldots ,m_n} {x^{-m_2}\over (1-x)}
{y_1^{-m_3}\over (1-y_1)}\bigl({y_1\over 1-y_1}\bigr)^2 {m_2+m_4
\atopwithdelims ()
2m_3}\ldots {m_{n-2}+m_n \atopwithdelims () 2m_{n-1}}{m_{n-1}
\atopwithdelims () 2m_{n}}\cr
&=\sum_{m_2,\ldots ,m_n}{1\over (1-x)}{y_1^{-m_3}\over (1-y_1)}
y_2^{m_2} {m_2+m_4
\atopwithdelims ()
2m_3}\ldots {m_{n-2}+m_n \atopwithdelims () 2m_{n-1}}{m_{n-1}
\atopwithdelims () 2m_{n}},}}
where
\eqn\defytwo{y_2\equiv \bigl({1\over\sqrt x}{y_1\over
(1-y_1)}\bigr)^2.}
The pattern is thus evident and the process is repeated until we end
up with
\eqn\endsum{f(x)=\prod_{j=0}^n {1\over (1-y_j)},}
where
\eqn\defys{y_{-1}\equiv 1, y_0=x, \hbox{\hskip 10pt \rm and\hskip 10pt}
(y_j^{-1} - 1)^2 = y_{j-1}^{-1}y_{j+1}^{-1}.}

These are the same recurrence relations as before, only this time,
\eqn\yj{y_j^{-1/2}=U_{j+1}({1\over{2\sqrt x}})}.

At this point it is easy to see that if we were to consider any other
representation, not necessarily $s=1$, there would be a term involving
$M$ in the binomial coefficient at a different place, the
$s^{\hbox{\rm th}}$
place. That is there would be a factor of the form

$${m_{s-1}+m_{s+1}+{\cal N} \atopwithdelims () 2m_{s}},$$
(where ${\cal N}$ is really equal to $N/2$), which would give rise to
a power $(y_{s-1})^{-\cal N}$ picked up while redifining the summation
variable so that the required sum is simply the constant term
(coefficient of $x^0$) in

$$f(x)=U_1({1\over{2\sqrt x}})U_s^{2 \cal N}({1\over{2\sqrt x}})
{U_{n+1}({1\over{2\sqrt x}}) \over U_{n+2}({1\over{2\sqrt x}})}.$$

Once again, using \ratcheb and closing the contours over the zeroes of
the denominator we arrive at
\eqn\finalsutwo{S_N={2\over l+2}\sum_{j=1}^{l+1} \sin^2 ({\pi j\over
l+2})U_s^N \bigl(\cos({\pi j\over l+2})\bigr),}
with $l=n+1$. This specializes correctly to the $s=1$ case evaluated
above.

As is immediately obvious, this method can be used for evaluating the
sums for cases where the quadratic form $K_{i j}^{a b}$ involves only
the $i, j$ or the $a,b$ indices.  The other cases require the
introduction of multiple variable generating functions and
multidimensional residues are required. For more results and techniques
associated with counting solutions of Bethe equations, see
\rKircount and \rComb.

\end